1

# A 16-Year Photometric Campaign on the Eclipsing Novalike Variable DW Ursae Majoris

D. R. S. Boyd[1], E. de Miguel[2,3], J. Patterson[4], M. A. Wood[5], D. Barrett[6], J. Boardman[7], O. Brettman[8], D. Cejudo[9], D. Collins[10], L. M. Cook[11], M. J. Cook[12], J. L. Foote[13], R. Fried[14], T. L. Gomez[15,16], F.-J. Hambsch[17], J. L. Jones[18], J. Kemp[19], R. Koff[20], M. Koppelman[21], T. Krajci[22], D. Lemay[23], B. Martin[24], J. V. McClusky[25], K. Menzies[26], D. Messier[27], G. Roberts[28], J. Robertson[29], J. Rock[30], R. Sabo[31], D. Skillman[32], J. Ulowetz[33], T. Vanmunster[34]

[1] *CBA Oxford, West Challow Observatory, 5 Silver Lane, West Challow, OX12 9TX, UK*
[2] *Departamento de Ciencias Integradas, Facultad de Ciencias Experimentales, Universidad de Huelva, 21071 Huelva, Spain*
[3] *CBA Huelva, Observatorio del CIECEM, Parque Dunar, Matalascañas, 21760 Almonte, Huelva, Spain*
[4] *Department of Astronomy, Columbia University, 550 West 120th Street, New York, NY 10027, USA*
[5] *Department of Physics & Astronomy, Texas A&M University-Commerce, Commerce, TX 75429, USA*
[6] *Observatory Le Marouzeau, 6 Le Marouzeau, St Leger Bridereix F-23300, France*
[7] *CBA Wisconsin, Luckydog Observatory, 65027 Howath Road, de Soto, WI 54624, USA*
[8] *CBA Huntley, 13915 Hemmingsen Road, Huntley, IL 60142, USA*
[9] *CBA Madrid, Observatorio El gallinero, El Berrueco, 28192 Madrid, Spain*
[10] *College View Observatory, 138 College View Drive, Swannanoa, NC 28778, USA*
[11] *CBA Concord, 1730 Helix Court, Concord, CA 94518, USA*
[12] *CBA Ontario, Newcastle Observatory, 9 Laking Drive, Newcastle, ON L1B 1M5, Canada*
[13] *CBA Utah, 4175 E. Red Cliffs Drive, Kanab, UT 84741, USA*
[14] *Deceased, formerly at Braeside Observatory, Flagstaff, AZ 86002, USA*
[15] *Instituto de Ciencias Matemáticas, Campus de Cantoblanco UAM, 28049 Madrid, Spain*
[16] *CBA Cuenca, 16540 Bonilla, Cuenca, Spain*
[17] *CBA Mol, Andromeda Observatory, Oude Bleken 12, B-2400 Mol, Belgium*
[18] *CBA Oregon, 22665 Bents Road NE, Aurora, OR 97002, USA*
[19] *Department of Physics, Middlebury College, Middlebury, VT 05753, USA*
[20] *CBA Colorado, Antelope Hills Observatory, 980 Antelope Drive West, Bennett, CO 80102, USA*
[21] *6019 Fairwood Drive, Minnetonka, MN 55345, USA*
[22] *CBA New Mexico, PO Box 1351 Cloudcroft, NM 88317, USA*
[23] *CBA Quebec, 195 Rang 4 Ouest, St-Anaclet, QC, G0K 1H0, Canada*
[24] *CBA Alberta, King's University College, Department of Physics, 9125 50th Street, Edmonton, AB T5H 2M1, Canada*
[25] *Blackburn College, 700 College Ave., Carlinville, IL 62626, USA*
[26] *CBA Massachusetts, 318A Potter Road, Framingham, MA 01701, USA*
[27] *CBA Norwich, 35 Sergeants Way, Lisbon, CT 06351, USA*
[28] *CBA Tennessee, 2007 Cedarmont Drive, Franklin, TN 37067, USA*
[29] *Arkansas Tech University, Department of Physical Science, 1701 North Boulder Avenue, Russellville, AR 72801, USA*
[30] *CBA Wilts, 2 Spa Close, Highworth, Swindon, SN6 7PJ, UK*
[31] *CBA Montana, 2336 Trailcrest Drive, Bozeman, MT 59718, USA*
[32] *CBA Mountain Meadows, 6-G Ridge Road, Greenbelt, MD 20770, USA*
[33] *CBA Illinois, Northbrook Meadow Observatory, 855 Fair Lane, Northbrook, IL 60062, USA*
[34] *CBA Belgium, Walhostraat 1A, B-3401 Landen, Belgium*

**ABSTRACT**

We present an analysis of photometric observations of the eclipsing novalike variable DW UMa made by the CBA consortium between 1999 and 2015. Analysis of 372 new and 260 previously published eclipse timings reveals a 13.6 year period or quasi-period in the times of minimum light. The seasonal light curves show a complex spectrum of periodic signals: both positive and negative "superhumps", likely arising from a prograde apsidal precession and a retrograde nodal precession of the accretion disc. These signals appear most prominently and famously as sidebands of the orbital frequency; but the precession frequencies themselves, at 0.40 and 0.22 cycles per day, are also seen directly in the power spectrum. The superhumps are sometimes seen together, and sometimes separately. The depth, width and skew of eclipses are all modulated in phase with both nodal and apsidal precession of the tilted and eccentric accretion disc. The superhumps, or more correctly the precessional motions which produce them, may be essential to understanding the mysterious "SW Sextantis" syndrome. Disc wobble and eccentricity can both produce Doppler signatures inconsistent with the true dynamical motions in the binary, and disc wobble might boost the mass-transfer rate by enabling the hot white dwarf to directly irradiate the secondary star.

## 1 INTRODUCTION

DW Ursae Majoris is an eclipsing novalike cataclysmic variable (CV) in which matter is transferring via Roche lobe overflow from a cool main sequence (MS) secondary star to a hot white dwarf (WD) primary via an optically thick accretion disc. In novalike variables this mass transfer is sustained at a high level, keeping the accretion disc in a bright state and avoiding the thermal instability which leads to dwarf nova outbursts. The novalikes also occasionally experience low states, when mass transfer is interrupted and the star remains very faint until the normal high mass-transfer rate is restored.

DW UMa is also a founding member of the informal, observationally-defined sub-class of SW Sextantis stars (Thorstensen et al. 1991). These are CVs which share most, but rarely all, of the following properties: high inclination, very high accretion rate, V-shaped eclipses, orbital period in the range 3-4 hours, positive and/or negative superhumps, and single-peaked emission lines highly out of phase with the true dynamical motions in the binary (Thorstensen et al. 1991; Rodriguez-Gil et al. 2007; Schmidtobreick, Rodriguez-Gil, & Gaensicke 2012).

This suite of behaviour, along with the possible significance of an evolutionary phase which must be brief (because the accretion rate is so high), has made SW Sex stars popular targets for study. But there is as yet no comprehensive explanation of SW Sex behaviour. Different studies adopt slightly different definitions of the class, and it is not yet clear whether the observed diversity comes from the stars, or merely from the astronomers who study them. But if a Rosetta Stone is ever found, it is likely to be DW UMa. This star is one of the brightest class members, and it exhibits deep eclipses in both high- and low-accretion states. These eclipses allow good luminosity and geometrical constraints for the several components of the binary: secondary, white dwarf, accretion disc, and even accretion-disc rim (Knigge et al. 2000, 2004; Dhillon, Smith & Marsh 2013).

Previous photometric studies of DW UMa have been published by Shafter, Hessman & Zhang (1988), Honeycutt, Livio & Robertson (1993), Dhillon, Jones & Marsh (1994), Knigge et al. (2000), Biro (2000 & 2002), Araujo-Betancor et al. (2003), Stanishev et al. (2004), Patterson et al. (2005), Boyd & Gänsicke (2009), Hoard et al (2010), and Dhillon et al. (2013). These have established the basic parameters of the binary: a mid-M secondary star of mass 0.25(5) $M_\odot$ transferring matter at $\sim 10^{-8}$ $M_\odot$/year to a 0.9(1) $M_\odot$ WD via an accretion disc (likely tilted, to account for the negative superhumps, which may signify the disc's retrograde wobble). The orbital period is 0.136606527(3) d, typical of the SW Sex class.

But these studies have certainly not exhausted the bounty available from photometry. There have been several reports of positive and negative superhumps, but with many details still missing: their full Fourier spectrum; how the signals interact with each other and with the basic orbital clock; their degree of stability; and how the periods change over the years. Such observations probably carry important information about the organized motions of the disc, in particular its shape and perturbations. The orbital period itself, defined by the interval between eclipse minima, may itself be (and is) variable on long timescales. These matters require study with an intensive program of time-series photometry over a very long baseline. We present such a study here, comprising 1342 hours of coverage during the years 1999-2015.

## 2 OBSERVATIONS AND DATA REDUCTION

The observations were made by the Centre for Backyard Astrophysics (CBA) consortium in 14 of the 17 observing seasons between 1999 and 2015. The CBA is a distributed community of mostly amateur astronomers who observe CVs on a long-term basis in order to understand the subtleties of their behaviour. Our extended campaign involved 31 observers in 6 countries distributed across North America and Europe, typically using 0.2 to 0.4m aperture telescopes equipped with CCD cameras. Observations were generally made with a clear (*C*) filter to maximise signal-to-noise while affording good resolution of time-dependent behaviour. Exposures were usually between 20 and 60 seconds depending on instrumentation and observing conditions. In 1999, when DW UMa was in a low state, observations were made with the 1.3m and 2.4m MDM telescopes at Kitt Peak using *C*, *B*, *Rc* and *Ic* filters. Table 1 gives an annual log of observations.

**Table 1.** Log of observations and measured eclipses.

| Year | No of runs | No of images | No of eclipses | Total time (h) |
|---|---|---|---|---|
| 1999 | 11 | 1267 | 6 | 30.66 |
| 2000 | 17 | 5530 | 25 | 93.24 |
| 2001 | 23 | 4975 | 25 | 84.70 |
| 2002 | 19 | 6761 | 33 | 124.66 |
| 2004 | 38 | 11787 | 51 | 215.54 |
| 2007 | 4 | 196 | 3 | 3.35 |
| 2008 | 6 | 1962 | 8 | 19.42 |
| 2009 | 2 | 142 | 2 | 2.54 |
| 2010 | 4 | 306 | 4 | 2.32 |
| 2011 | 10 | 2678 | 10 | 34.36 |
| 2012 | 19 | 4106 | 26 | 85.13 |
| 2013 | 9 | 2226 | 9 | 22.81 |
| 2014 | 78 | 29162 | 114 | 429.65 |
| 2015 | 30 | 13079 | 56 | 192.95 |
| Total | 270 | 84177 | 372 | 1341.33 |

Observers dark-subtracted and flat-fielded their images before performing differential aperture photometry of DW UMa with respect to a nearby comparison star, typically one chosen from the AAVSO comparison chart (AAVSO 2015) with as close a colour index to DW UMa as possible. Magnitudes of comparison stars from AAVSO field photometry were used to derive instrumental *C* magnitudes for DW UMa. *B*, *Rc* and *Ic* magnitudes for the magnitude 17 comparison star used in 1999 were calculated from Sloan *griz* magnitudes (SDSS Data Release 6 2015) and used to derive *B*, *Rc* and *Ic* magnitudes for DW UMa in 1999. As the QE of the CCD camera used at Kitt Peak peaked in the *Rc* passband, the comparison star *Rc* magnitude was used to obtain the *C* magnitude of DW UMa. Observers typically provided a Julian Date (JD) and magnitude of DW UMa for each image processed. Prior to analysis, all times were converted to Heliocentric JDs.

Because of the different spectral responses of the equipment used and different choice of comparison stars, each observer's data were on a slightly different instrumental *C* magnitude scale. In general, but not always, observers were consistent in their choice of comparison star within each year and from year to year. The challenge was to find the small magnitude offsets for each observer's data, which would produce a single *C* magnitude scale. The data from one observer which spanned three years using the same comparison star were taken as defining a reference *C* magnitude to which all other observers' data were aligned in magnitude by adding or subtracting a magnitude offset. Where datasets overlapped in time, this alignment was performed visually. Where they did not overlap, for example between years, the reference level was propagated using the data of observers who consistently used the same equipment and comparison stars. The accuracy of aligning overlapping datasets was checked analytically for a sample of data which showed that it produced alignment to better than 0.02 magnitudes (Lloyd, private communication). In other cases, the mean alignment accuracy was 0.03-0.04 magnitudes. A composite C magnitude light curve for all years from 1999 to 2015 is shown in Fig. 1.

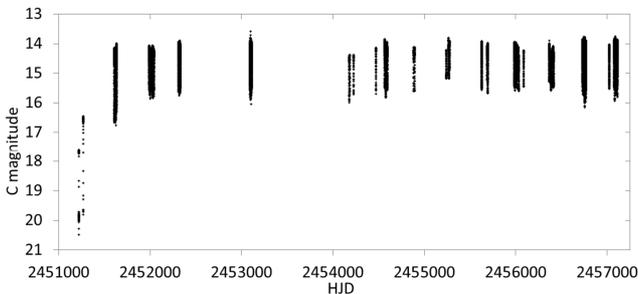

**Figure 1**. C magnitude light curve for all data from 1999 to 2015.

Using one observer's equipment we measured the standard Johnson V magnitude of DW UMa using *B* and *V* filters and at the same time its *C* magnitude. This gave us the relationship between that observer's instrumental *C* magnitude and the *V* magnitude standard. We then adjusted this for the offset between that observer's *C* magnitude and our reference *C* magnitude described above. This gave the following relationship between our reference instrumental *C* magnitude and a standard *V* magnitude

$V = C - 0.07 \pm 0.04$ (1)

The individual light curves were initially phased on the orbital period given in Stanishev et al. (2004) in order to identify the location of eclipses and to determine a cycle number for each eclipse. In Fig. 2 we show, with a consistent magnitude scale, mean orbital light curves recorded with a clear filter for those years between 2000 and 2015 for which we have full coverage in orbital phase. In other years we concentrated only on the eclipse region, in order to measure eclipse timings. A gradual recovery from the low state is apparent in the early years. In the high state we see eclipses only of the accretion disc, but with sometimes a shallow dip around phase 0.5. This could possibly be a secondary eclipse, although other interpretations are possible (Dhillon et al. 2013). There are other small bumps and dips which could be significant, but are too weak or transient to assess. Lengthy coverage at other wavelengths, especially the infrared (sampling the secondary star and outer disc), might clarify this matter.

## 3 ECLIPSE TIMING AND ORBITAL PERIOD

The eclipse segments of light curves (between phases -0.15 and +0.15) were extracted for analysis of eclipse timing. For the broad V-shaped eclipses seen in 2000 to 2015, times of minimum light were obtained by fitting a quadratic polynomial to the lower part of each eclipse. Typically this was the lower third of the eclipse but with discretion applied where the shape of the light curve within the eclipse was distorted. The eclipses in 1999 were steep-sided and flat-bottomed as we shall see in section 8. The mid time of the flat bottom of the eclipse obtained from linear fits to the ingress, bottom and egress was taken as the time of minimum.

In total, 372 eclipses were measured, and are listed by year in Table 1. Their cycle numbers and times of minimum are given in Table 2.

**Table 2**. Times of minimum, scaled errors and cycle numbers of 372 measured eclipses. The full table is given in the Appendix.

| HJD | Error | Cycle |
|---|---|---|
| 2451210.77362 | 0.00040 | 0 |
| 2451217.05719 | 0.00060 | 46 |
| 2451218.01388 | 0.00040 | 53 |
| 2451218.83332 | 0.00040 | 59 |
| 2451218.97010 | 0.00060 | 60 |
| 2451219.78935 | 0.00060 | 66 |

A linear ephemeris for these data was calculated as

$T_{min}$ (HJD) = 2451605.97629(2) + 0.1366065346(8) *E*. (2)

An *O-C* (observed minus calculated) plot of the residuals with respect to this linear ephemeris (Fig. 3) appeared to show a progressive decrease in the orbital period between 1999 and 2015. Over this period a quadratic ephemeris fitted the data better than a linear ephemeris.

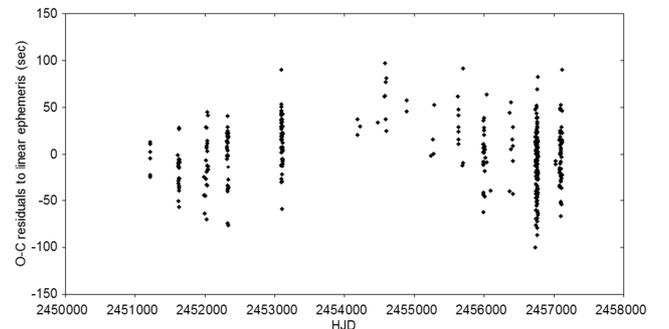

**Figure 3**. *O-C* residuals of eclipse timings with respect to the linear ephemeris in equation (2) for 372 CBA eclipses.

The polynomial fits to the eclipse minima provided statistical errors on the times of minimum which gave an indication of their relative precision but were underestimates of the true errors. Erratic fluctuations in brightness (flickering) and changes of the accretion disc's shape certainly increase the errors.

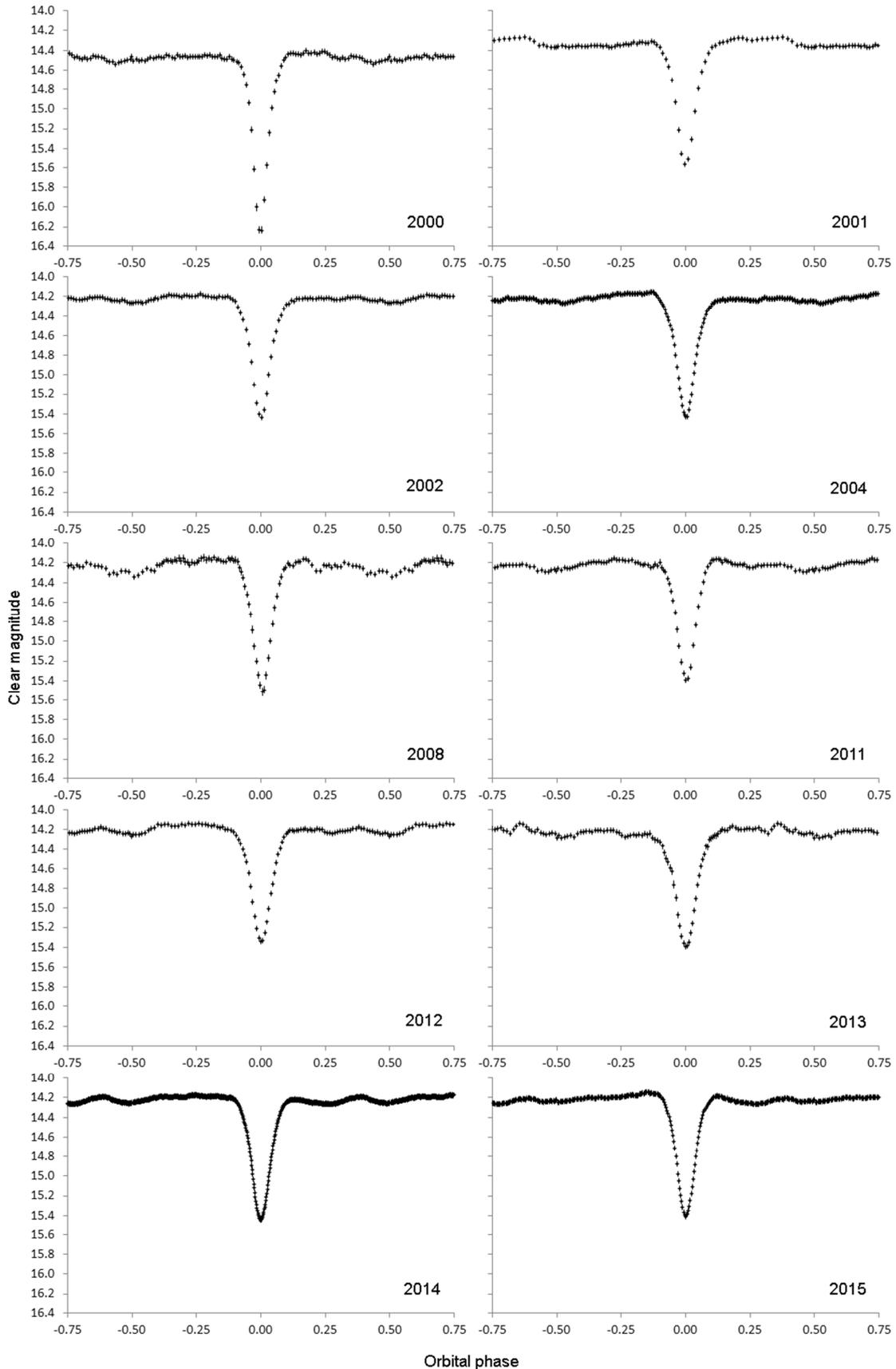

**Figure 2**. Mean orbital light curves for years in which we have sufficient data. The gradual recovery from the 1999 low state is visible, and there are often small secondary dips around phase 0.5.

To determine realistic errors on the measured times of minimum, a single scaling factor was applied to all the errors calculated from the polynomial fits to make the root mean square (rms) of these scaled errors match the rms *O-C* residual of the times with respect to the quadratic ephemeris which we assumed to be a good representation of the data over this long interval. In this way we preserved the relative sizes of the errors in the eclipse timings while ensuring that they were consistent with the real scatter in the data. These scaled errors are also included in Table 2.

We found many additional eclipse timings from published papers dating back to 1983. High-quality timings came from Shafter et al. (1988), Dhillon et al. (1994), Biro (2000), Stanishev et al. (2004), Boyd & Gänsicke (2009), and Dhillon et al. (2013). Others, of lower or unknown quality, came from various issues of IBVS and BVSOLJ. This amounted to 260 additional timings. When we included them in the analysis, the picture changed, and a quadratic ephemeris was clearly no longer viable. To determine an ephemeris we needed the errors of these timings, but in many cases the published errors were clearly too small. To estimate more reliable errors, a separate linear ephemeris was calculated using only the data in each published study. The real scatter of the eclipse timings with respect to that linear ephemeris, as measured by their rms *O-C* residual, was taken as a more reliable value for the error for all eclipse timings given in that study. We consider this a valid approach as the duration of each study covered only a small time interval.

This gave us a total of 632 eclipse timings, with what we consider to be realistic estimates of their errors. Weighting each time by the inverse square of its corresponding error, we calculated an improved linear ephemeris for DW UMa as follows:

$T_{min}$ (HJD) = 2451605.97651(2) + 0.1366065324(7) $E$     (3)

The corresponding mean orbital frequency is 7.32029415(4) c d$^{-1}$. *O-C* residuals of these 632 eclipse timings with respect to this linear ephemeris are shown in Fig. 4 (upper). Fig. 4 (middle) shows the annual mean and standard deviation of the *O-C* residuals for each year with at least two eclipse timings. There is a clear indication of sinusoidal behaviour. A weighted sine fit to these annual *O-C* residuals using PERIOD04 (Lenz & Breger 2005) is superimposed. The period and semi-amplitude are 13.6±0.4 yr and 25.7±3.1 s respectively, and the fit has $\chi^2$ of 5.6 for 18 degrees of freedom. The annual *O-C* residuals with respect to this sine fit are shown in Fig. 4 (lower) and have an rms of 15.0 s.

## 4 ECLIPSE TIMING VARIATION

With so many eclipses included – vastly more than any other CV with similar wiggles in the *O-C* – it is a reasonable hypothesis that the wiggles of DW UMa signify true variations in the orbital period. Changes in the shape of the eclipsed object (the disc) must certainly modify the individual timings and this doubtless contributes to the scatter seen in the upper diagram in Fig. 4, but such effects must be on much shorter timescales. The two most widely proposed explanations for a periodic decades-long term in the *O-C* are: (a) a third body in the system which introduces a variable delay as the DW UMa binary centre of mass moves around the common barycentre of the triple system, and (b) changes in orbital period because of changes in the secondary's mass distribution, due to cyclic changes in its magnetic field, the so-called Applegate mechanism (Applegate 1992).

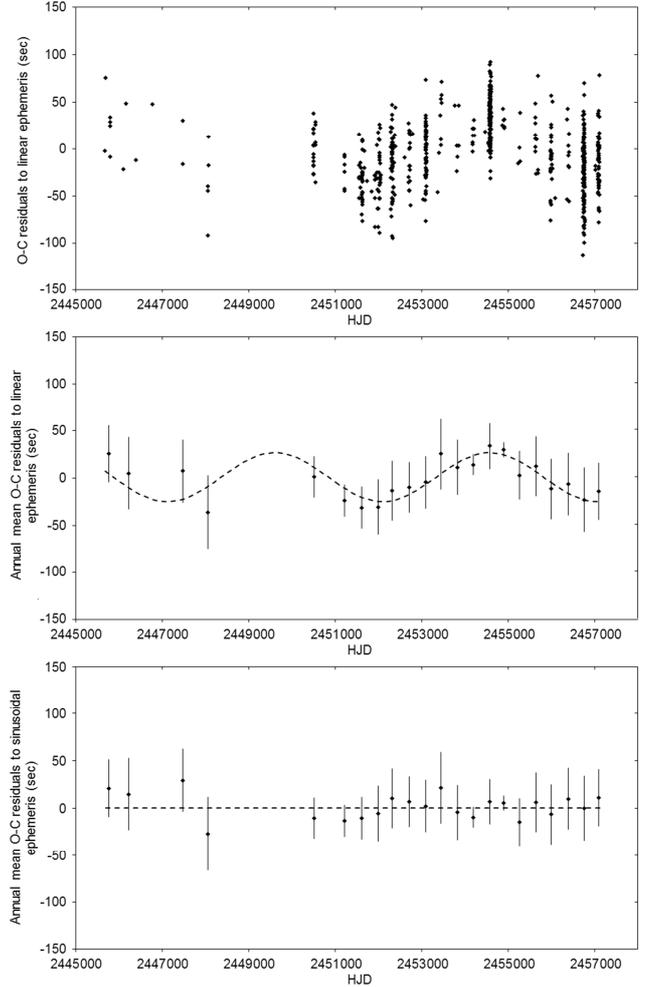

**Figure 4**. *O-C* residuals of eclipse timings with respect to the linear ephemeris in equation (3) for 632 eclipses (upper), annual mean *O-C* residuals with respect to this linear ephemeris showing a fitted sinusoidal ephemeris (middle) and annual mean *O-C* residuals with respect to the sinusoidal ephemeris (lower).

To investigate the third body hypothesis we make the assumption, consistent with the data shown in Fig. 4, that the eclipse timing variation is sinusoidal and therefore the orbit of the DW UMa binary centre of mass around the common barycentre is circular. From Hilditch (2001) we have $a_1 \sin i = A\,c = 0.052$ AU and mass function

$$\begin{aligned} f(m_2) &= m_2^3 \sin^3 i / (m_1 + m_2)^2 \\ &= (2\pi)^2 A^3 c^3 / (G\,P^2) \\ &= 7.39 \times 10^{-7}\,M_\odot \end{aligned} \quad (4)$$

where $m_1$ and $m_2$ are the masses of the DW UMa binary system and the third body, $a_1$ is the orbital radius of the DW UMa centre of mass about the common barycentre, $i$ is the inclination of the third body orbit, and $A$ and $P$ are the semi-amplitude and period of the sinusoidal variation in the eclipse *O-C* plot. From equation (4) we have the relationship between the mass of the third body and its orbital inclination to our line of sight.

This is shown in Fig. 5 (left), while Fig. 5 (right) shows the relationship between the radius of the third body orbit about the common barycentre and its inclination. If we assume this third body to be the sole source of the *O-C* modulation, and its orbit to be coplanar with the DW UMa binary orbit, then its mass would be 10.06 $M_{Jup}$ and its orbital radius 5.80 AU. Therefore this body, if it exists, is almost certainly planetary not stellar.

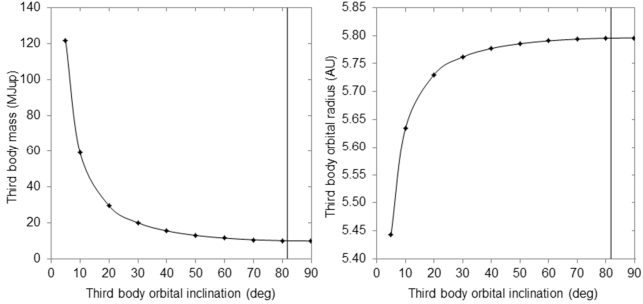

**Figure 5**. Variation of mass (left) and orbital radius (right) with orbital inclination for a possible third body. The orbital inclination of DW UMa reported by Araujo-Betancor (2003) is marked by a vertical line.

The Applegate mechanism is a potential cause of eclipse timing modulation in binaries containing a star with an active magnetic cycle. According to the formulation of the Applegate mechanism derived by Watson & Marsh (2010) (their equation 13), the potential *O-C* variation $\delta t$ in seconds is given by

$$\delta t \leq k \, (\Omega/\Omega_\odot) \, (M/M_\odot)^{-3/2} \, (R/R_\odot)^4 \, (L/L_\odot)^{1/2} \, a^{-2} \, T^{3/2} \qquad (5)$$

where $\Omega$ is the angular velocity, $M$ the mass, $R$ the radius and $L$ the luminosity of the MS star, $a$ is the DW UMa binary separation in AU, $T$ is the time-scale over which the magnetic changes occur in years. The quantity $k = 2.172 \times 10^{-4} \, \alpha^{1/2} \, \beta^{1/2}$ where $\alpha$ is the fractional upper limit of luminosity variations of the MS star, and $\beta$ is the fraction of the mass of the MS star in the outer shell which experiences differential rotation relative to the stellar core.

We take the DW UMa binary system parameters from Araujo-Betancor et al. (2003), the temperature of the MS star as ~3300K from our spectral type of M3.1 found below in section 8, and *T* as 13.6±0.4 yr, the observed *O-C* modulation period. We assume that rotation of the MS star is tidally locked to its binary orbital period. Estimating the value of $\alpha$ is difficult because we cannot be absolutely certain of the alignment of our magnitude measurements from year to year. Nevertheless we conservatively believe we would have detected a systematic magnitude variation over the putative magnetic cycle of 0.2 magnitudes or larger. This gives a value for $\alpha$ of 0.17. According to Applegate (1992), a reasonable value for $\beta$ is 0.1. We assume errors of 10 per cent in $\alpha$ and $\beta$.

On this basis the upper limit for $\delta t$ is 85±67 s with the error propagated from the errors on the parameters in equation (5). We therefore conclude that the Applegate mechanism is a possible source of the observed 13-year variation in the *O-C*.

Only further long term study of eclipse timings in DW UMa will establish whether they follow a strict period, which would favour a third body interpretation, or turn out to be only quasi-periodic favouring some other explanation, which might be related to a magnetic periodicity in the MS star.

## 5 PERIODIC SIGNALS IN THE LIGHT CURVE OUTSIDE THE ECLIPSES

Of course, the principal features in the light curve are the eclipses which occur at the orbital frequency $\omega$. To look for subtler periodic signals, we remove the eclipses between phases -0.15 and +0.15 and study the seasonal (usually) light curves outside eclipse. We use the Date-Compensated Discrete Fourier Transform (DCDFT) technique in conjunction with the CLEANest algorithm (Foster 1995) implemented in PERANSO (Paunzen & Vanmunster 2016). This is particularly suitable for analysing irregularly spaced data with large gaps, and is effective in removing spurious or alias signals caused by the inevitable sampling window of the data, usually ±1 c d$^{-1}$ (cycles per day). It produces the set of single frequency signals and their amplitudes which best represent the data, plus a residual spectrum. The uncertainties in signal frequencies and amplitudes are calculated according to the method in Schwarzenberg-Czerny (1991). This uses the autocorrelation function of the residuals to determine their mean correlation length, which is used to scale the conventionally calculated correlation matrix to get the true correlation matrix.

To interpret the results from this analysis, we draw on previous experience of interpreting periodic signals in CVs (see for example the Appendix in Patterson et al. (2002a)). Periodic signals are sometimes observed in novalikes with a frequency slightly displaced from the orbital frequency $\omega$. These are the famous "superhumps". Since the studies of the early 1990s (Patterson et al. 1993, Harvey et al. 1995), it has been generally accepted that these arise from perturbations of the accretion disc resulting in disc precession. Observations show that apsidal precession is always prograde (with frequency denoted $\Omega$) and nodal precession is always retrograde (with frequency denoted $N$). The resultant periodic signals mainly occur at the difference frequencies $\omega-\Omega$ (positive superhumps) and $\omega+N$ (negative superhumps), although sometimes the nodal precession frequency $N$ appears in the power spectrum. The theoretical understanding for these photometric waves was launched by the simulations of Whitehurst (1988), and has now reached a level of detail almost great enough to confront the complexity of the observations (Wood, Thomas & Simpson 2009; Montgomery 2012a, 2012b). .

DW UMa is rich in superhumps. Fig. 6 shows a DCDFT power spectrum of the light curve outside eclipse for 2014. The most prominent signals are at 0.218 c d$^{-1}$ and 7.540 c d$^{-1}$. There are also ±1 c d$^{-1}$ aliases of these signals, plus another signal (also an alias) at (1-0.218) c d$^{-1}$. As the two most prominent signals differ by the orbital frequency, and the period of the 7.540 c d$^{-1}$ signal is shorter than the orbital period, we interpret the higher frequency signal as being due to a negative superhump and the 0.218 c d$^{-1}$ signal as due to nodal precession. Also shown in Fig. 6 is the spectral window of the same data. As well as the expected peak at zero and its +1 c d$^{-1}$ alias, there are signals at the orbital frequency 7.320 c d$^{-1}$ and its first harmonic, caused by the gaps left by removal of the eclipses, and their ±1 c d$^{-1}$ aliases.

Examining CLEANest power spectra for all years we found no significant signals beyond 10 c d$^{-1}$ in any year, and power concentrated in the intervals 0–2 c d$^{-1}$ and 4–10 c d$^{-1}$. Only in 2014 and 2015 was there sufficient data to cleanly detect long period signals below 2 c d$^{-1}$. This was not surprising as these two years contained the most extensive data.

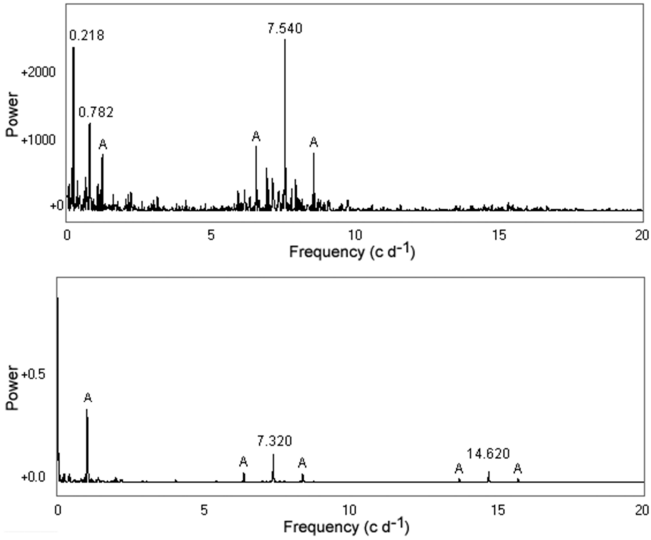

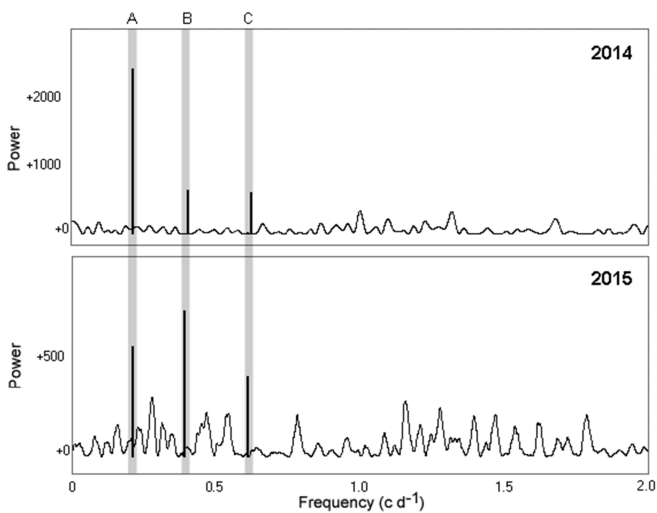

**Figure 6**. (Upper) DCDFT power spectrum of the light curve for 2014 outside eclipse. The strongest signals are at 0.218 c d$^{-1}$ (nodal precession) and 7.540 c d$^{-1}$ (negative superhumps). Also present are ±1 c d$^{-1}$ aliases of these signals (marked A) and another alias at (1-0.218) c d$^{-1}$. (Lower) Spectral window of the data with the expected peak at zero and its +1 c d$^{-1}$ alias plus signals at the orbital frequency 7.320 c d$^{-1}$, its first harmonic and their ±1 c d$^{-1}$ aliases.

Fig. 7 shows CLEANest power spectra in the frequency range 0–2 c d$^{-1}$ for 2014 and 2015 with three prominent signals labelled A, B and C. The CLEANest algorithm has effectively removed the alias signals seen in the window function in Fig. 6 from the power spectrum. We have already identified signal A around 0.22 c d$^{-1}$ as being due to nodal precession and we now speculate that signal B around 0.40 c d$^{-1}$ is due to apsidal precession. The frequency of signal C is the sum of the frequencies of A and B ($\Omega+N$). The frequencies and semi-amplitudes of these signals are listed in Table 3.

**Figure 7**. CLEANest power spectra for 2014 and 2015 in the frequency range 0–2 c d$^{-1}$. We interpret signals A and B as due to nodal and apsidal precession respectively. Signal C is the sum of frequencies A and B. The frequencies and semi-amplitudes of these signals are given in Table 3.

Fig. 8 shows CLEANest power spectra in the frequency range 4–10 c d$^{-1}$ for 2002, 2004, 2014 and 2015, the four years with sufficient data for analysis of this frequency range. We have highlighted five signals labelled P to T which are present in at least two of these four years. Their frequencies and semi-amplitudes are listed in Table 3.

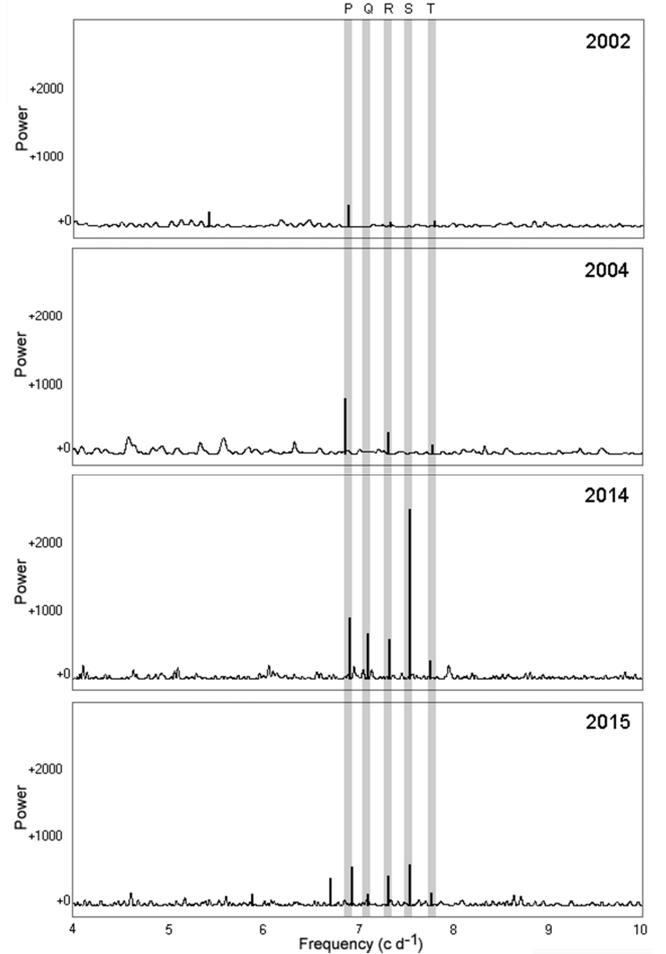

**Figure 8**. CLEANest power spectra for 2002, 2004, 2014 and 2015 in the frequency range 4–10 c d$^{-1}$. Five signals labelled P to T are each present in at least two years. Our interpretation of these signals is described in the text and their frequencies and semi-amplitudes are given in Table 3.

The signals labelled R are all close to the binary orbital frequency. We interpret signals P and S as being caused by positive and negative superhumps respectively. Their values are consistent with those reported in Patterson et al. (2002b), Stanishev et al. (2004) and Boyd & Gänsicke (2009). The difference in frequency between signals R and P, which we would expect to be the apsidal precession frequency, is approximately 0.40 c d$^{-1}$ thus strengthening our speculation that the signals we saw at that frequency in 2014 and 2015 are indeed due to apsidal precession. Signals Q and T correspond to frequencies $\omega-N$ and $\omega+\Omega$ about which we shall say more in section 7. There is also a signal at 6.7060(4) c d$^{-1}$ in 2015 which is close to the frequency $\omega-(\Omega+N)$ (6.6993 c d$^{-1}$).

**Table 3**. Frequencies and semi-amplitudes of signals from CLEANest analyses of the light curves outside eclipse in 2002, 2004, 2014 and 2015 and our interpretation of their origin. Figures in brackets are uncertainties in the final digit calculated according to the method in Schwarzenberg-Czerny (1991).

| Signal | | 2002 | 2004 | 2014 | 2015 | Interpretation |
|---|---|---|---|---|---|---|
| A | Frequency (c d$^{-1}$) | | | 0.2190(3) | 0.2174(3) | Nodal precession $N$ |
|   | Semi-amplitude (mag) | | | 0.097(2) | 0.067(3) | |
| B | Frequency (c d$^{-1}$) | | | 0.4092(7) | 0.3934(3) | Apsidal precession $\Omega$ |
|   | Semi-amplitude (mag) | | | 0.043(2) | 0.062(3) | |
| C | Frequency (c d$^{-1}$) | | | 0.6289(6) | 0.6134(3) | $\Omega + N$ |
|   | Semi-amplitude (mag) | | | 0.051(2) | 0.059(3) | |
| P | Frequency (c d$^{-1}$) | 6.880(2) | 6.864(2) | 6.9124(6) | 6.9292(3) | +ve superhump $\omega - \Omega$ |
|   | Semi-amplitude (mag) | 0.045(4) | 0.065(3) | 0.049(2) | 0.063(3) | |
| Q | Frequency (c d$^{-1}$) | | | 7.1020(7) | 7.0996(5) | $\omega - N$ |
|   | Semi-amplitude (mag) | | | 0.042(2) | 0.037(3) | |
| R | Frequency (c d$^{-1}$) | 7.327(4) | 7.317(2) | 7.3228(5) | 7.3084(3) | Binary orbit $\omega$ |
|   | Semi-amplitude (mag) | 0.026(4) | 0.040(3) | 0.053(2) | 0.053(3) | |
| S | Frequency (c d$^{-1}$) | | | 7.5409(3) | 7.5375(2) | -ve superhump $\omega + N$ |
|   | Semi-amplitude (mag) | | | 0.096(2) | 0.085(3) | |
| T | Frequency (c d$^{-1}$) | 7.788(4) | 7.780(3) | 7.7560(11) | 7.7596(5) | $\omega + \Omega$ |
|   | Semi-amplitude (mag) | 0.023(4) | 0.023(3) | 0.025(2) | 0.033(3) | |

The mean apsidal and nodal superhump period excesses are 0.063(4) and -0.029(1) respectively. These values are consistent with corresponding superhump period excesses for DW UMa of 0.064(2) and -0.029(2) reported in Patterson et al. (2002b). They are also broadly in agreement with the distributions of superhump period excess for positive and negative superhumpers given respectively in fig. 3 in Patterson (1998) and fig. 15 in Wood et al. (2009). The relationship

$$\varepsilon = 0.18\, q + 0.29\, q^2 \qquad (6)$$

in Patterson et al. (2005) where $\varepsilon$ is the apsidal superhump period excess and $q$ is the mass ratio, gives $q \sim 0.25$ which is at the low end of the range 0.39(12) given by Araujo-Betancor et al. (2003).

From the range of values found for the orbital frequency, which we know to be highly stable over this period, it is clear that the frequency uncertainties in Table 3 given by the CLEANest analysis are underestimates by a factor of approximately five. If we consider the implication of scaling the frequency uncertainties in Table 3 by a factor five, we conclude that the frequencies of nodal precision and its progeny (negative superhumps) are relatively stable from year to year while those of apsidal precession and positive superhumps are not.

Fig. 9 shows the light curve outside eclipse for 2014, the year in which we have most data, phased on the nodal and apsidal precession periods corresponding to the frequencies in Table 3. The sinusoidal curves have semi-amplitudes as given in Table 3. Phase zero corresponds to the maximum magnitude of the precession cycles at the epoch indicated. The large error bars reflect the spread caused by the superhumps at much higher frequencies.

Fig. 10 shows the light curves outside eclipse phased on the positive superhump periods corresponding to the frequencies in Table 3 for 2002, 2004, 2014 and 2015 and on the negative superhump periods for 2014 and 2015. In each case phase zero corresponds to the superhump maximum at the epoch indicated.

We now turn our attention to the likely apsidal precession signal around 0.40 c d$^{-1}$. We measured the frequency of this signal directly in 2014 and 2015 while in 2002, 2004, 2014 and 2015 we can find it indirectly as the difference between the frequency of the binary orbit from the orbital ephemeris and the positive superhump frequencies in Table 3. These values are tabulated in Table 4 with calculated average apsidal precession frequencies for each year and the corresponding periods. We have scaled the uncertainties from Table 3 by a factor five based on the discussion above. Also included in Table 4 are the frequency and period of the apsidal precession signal found in the 2008 campaign reported in Boyd & Gänsicke (2009). These are consistent with the results of this study. A full analysis of the data from the 2008 campaign will be reported in a separate paper.

Fig. 11 shows how the apsidal precession period has varied between 2002 and 2015 together with a speculative sinusoidal fit with a period of around 40 yr and semi-amplitude 0.36 d.

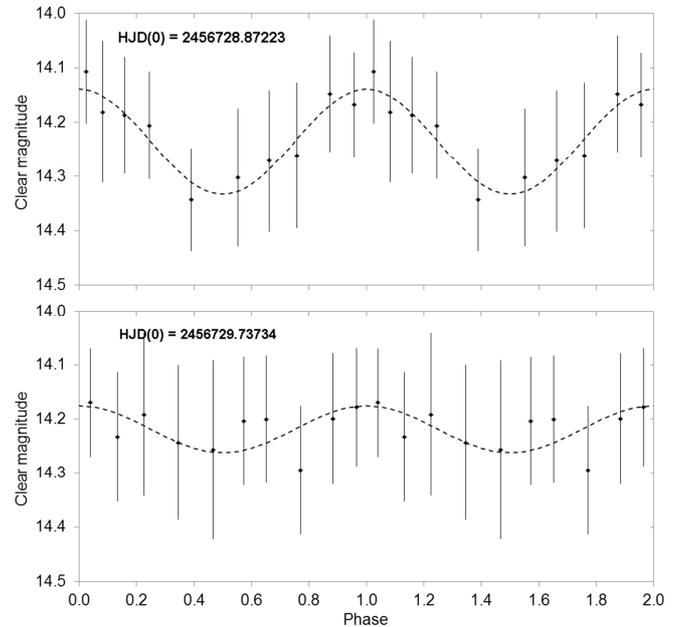

**Figure 9**. Light curve outside eclipse for 2014 phased at the nodal (upper) and apsidal (lower) precession periods corresponding to the frequencies in Table 3. The sinusoidal curves have semi-amplitudes as given in Table 3. Phase zero in both plots corresponds to the maximum magnitude of the precession cycles at the epoch shown. The large uncertainties are due to the spread in each bin caused by higher frequency superhumps. Two cycles are shown and bins are equally populated.

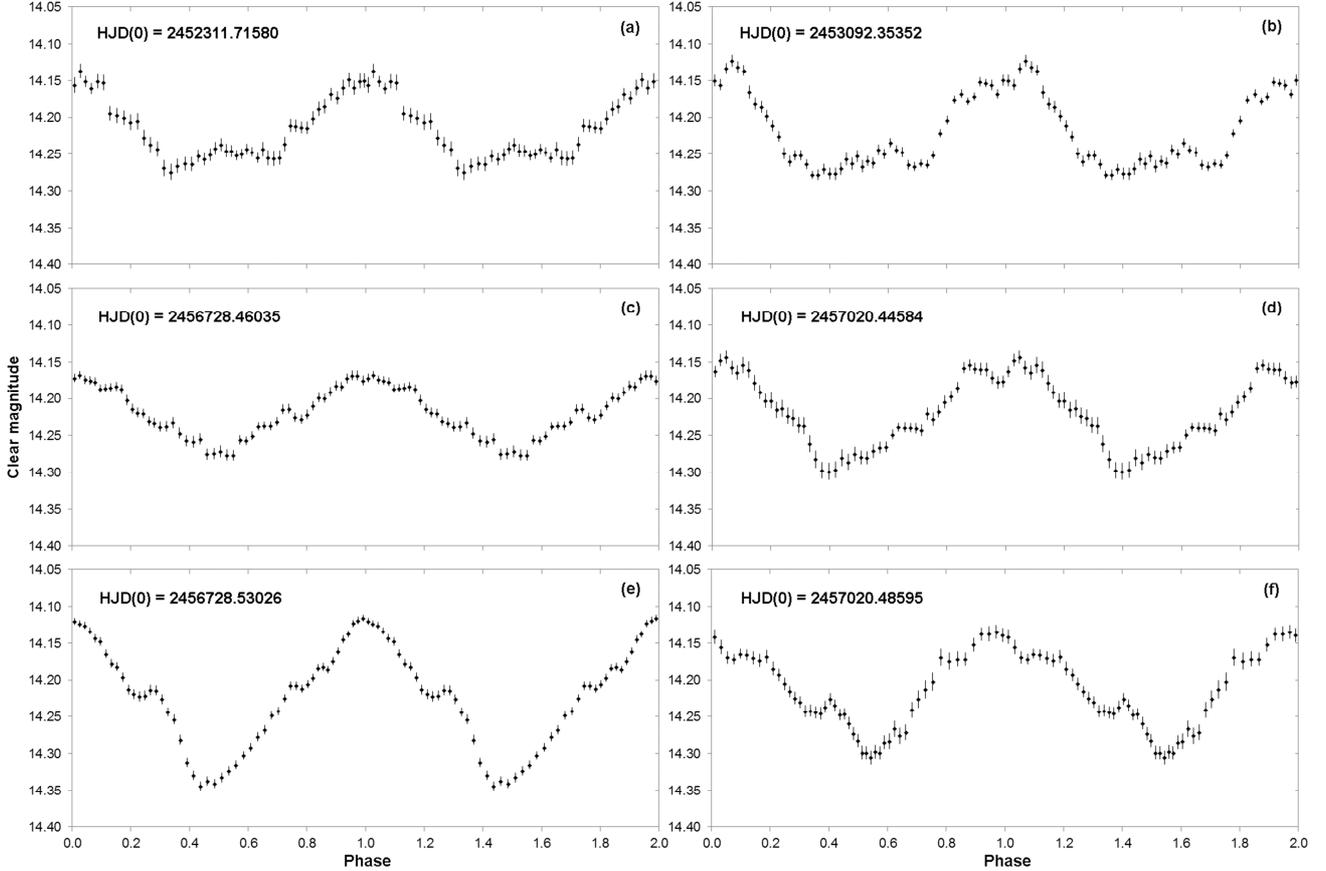

**Figure 10**. Light curves outside eclipse phased on the positive superhump periods for (a) 2002, (b) 2004, (c) 2014 and (d) 2015 and on the negative superhump periods for (e) 2014 and (f) 2015 corresponding to the frequencies in Table 3. In each case phase zero corresponds to the superhump maximum. Epochs of maximum are shown. Two cycles are shown and the vertical scale is the same in each case.

**Table 4**. Frequencies and periods for apsidal precession obtained in 2002, 2004, 2014 and 2015 from this study and in 2008 from the campaign reported in Boyd & Gänsicke (2009). Calculation of the uncertainties is described in the text.

| | 2002 | 2004 | 2008 | 2014 | 2015 |
|---|---|---|---|---|---|
| Apsidal precession frequency $\Omega$ (c d$^{-1}$) | | | | | |
| Measured directly | | | 0.448(1) | 0.409(4) | 0.393(2) |
| Calculated from 7.3203 c d$^{-1}$ - ($\omega$-$\Omega$) | 0.440(11) | 0.456(6) | 0.443(1) | 0.408(3) | 0.391(2) |
| Average | 0.440(11) | 0.456(6) | 0.446(3) | 0.409(5) | 0.392(2) |
| Apsidal precession period (d) | 2.27(5) | 2.19(3) | 2.24(2) | 2.45(3) | 2.55(1) |

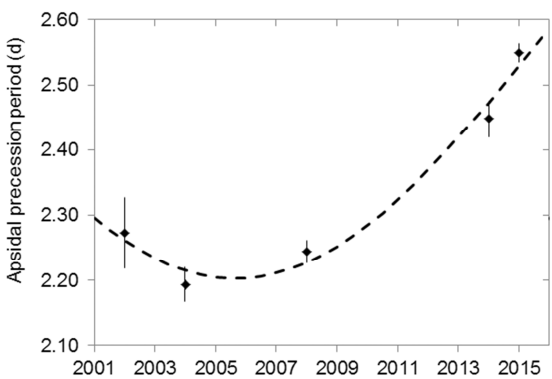

**Figure 11**. Variation of the apsidal precession period between 2002 and 2015. The data for 2002, 2004, 2014 and 2015 are from this study, the 2008 data comes from the campaign reported in Boyd & Gänsicke (2009). The dashed curve is a segment of a speculative sinusoidal fit.

## 6 ECLIPSE PROFILE VARIATION

We found the magnitudes before and after each eclipse by averaging the magnitudes of five data points before phase -0.15 and five points after phase +0.15. We then interpolated a magnitude between these at phase 0.0, and took this as representing the magnitude which the light curve would have had in the absence of the eclipse. The minimum magnitude of each eclipse was found from the quadratic fit to the eclipse. The difference between these magnitudes was taken as the eclipse depth. A depth was measured for 305 eclipses which were sufficiently well observed to enable this calculation.

The average of the interpolated magnitude of the light curve and the magnitude at minimum was taken as the half-depth magnitude for each eclipse. The difference in time between the ingress and egress sides of the eclipse at this half-depth magnitude was taken as the width of each eclipse. This width was measured for 301 sufficiently well observed eclipses.

Fig. 12 shows the mean eclipse depth and width for each year in which we have data between 2000 and 2015. Eclipses were noticeably deeper and narrower in 2000 while the disc was recovering from the low state.

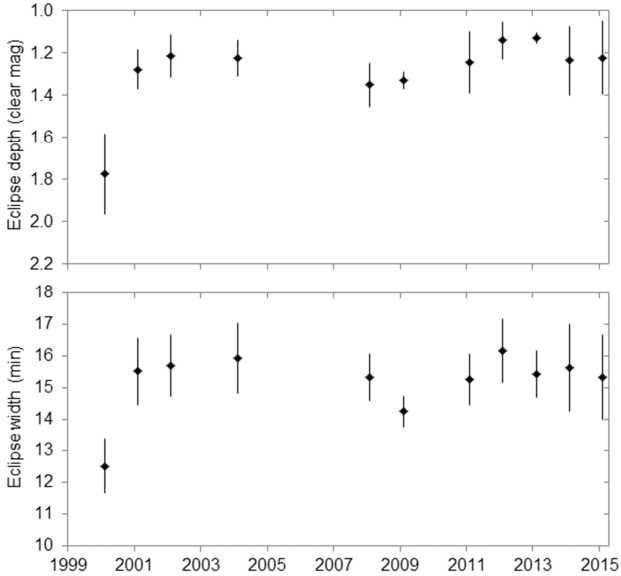

**Figure 12**. Mean eclipse depth and width, as defined in the text, for each year in which we have data between 2000 and 2015.

While examining the eclipses we noticed some were slightly skewed with different slopes on ingress and egress. To investigate this we calculated the centre time of each eclipse which we defined as the mid-point of the times of ingress and egress at the half-depth magnitude. Fig. 13 shows a histogram of the difference between the eclipse centre and eclipse minimum times for 301 eclipses. We refer to this time difference as the eclipse skew. A Kolmogorov–Smirnov test shows the distribution to be consistent with Gaussian at the 95 per cent confidence level. The mean time difference between eclipse centre and eclipse minimum is +7.17 s and, as the standard deviation of the mean is 1.56 s, this is 4.6 standard deviations from zero. We therefore see a statistically significant tendency for eclipses to be slightly skewed towards a delayed centre time through having steeper ingress and shallower egress.

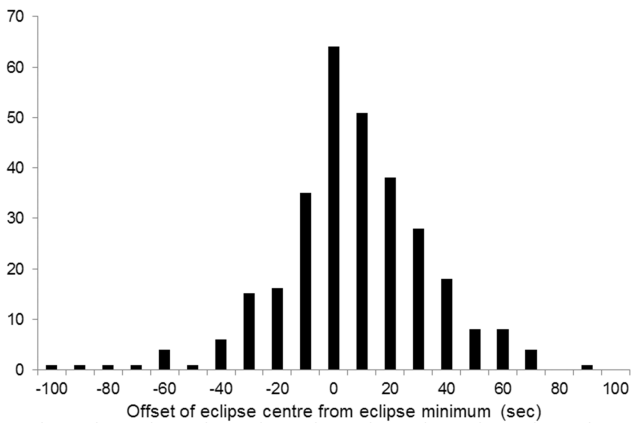

**Figure 13**. Histogram of the time difference between the eclipse centre and eclipse minimum (referred to in the text as eclipse skew).

It is also possible that the state of eccentricity and/or tilt of the nearly edge-on accretion disc at the time of each eclipse might affect the observed eclipse profile and therefore cause variation of the observed eclipse depth, width and skew. We investigated this by performing CLEANest frequency analyses of the variation of these three eclipse parameters with time. The results of these analyses for all sufficiently well observed eclipses between 2001 and 2015 are shown in Fig. 14.

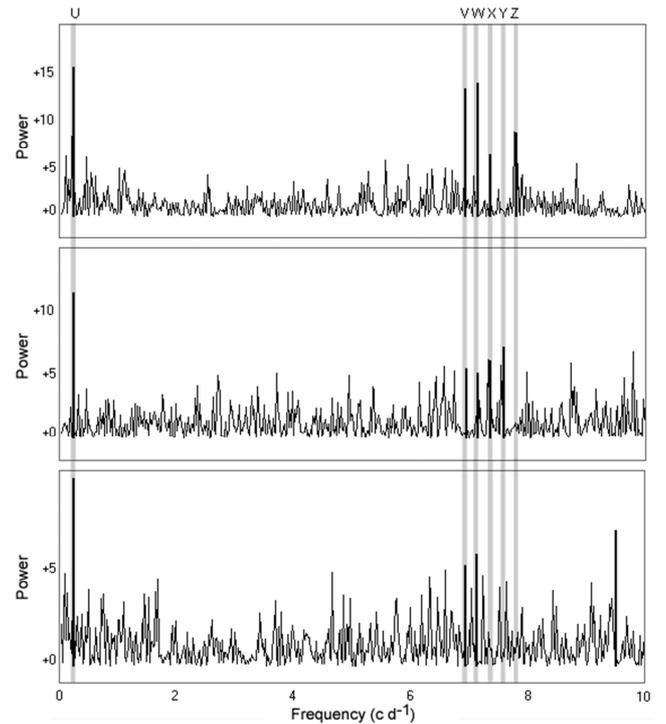

**Figure 14**. CLEANest power spectra of eclipse depth (upper), eclipse width (middle) and eclipse skew (lower) for all sufficiently well observed eclipses between 2001 and 2015. Six signals labelled U to Z are marked. Our interpretation of these signals is described in the text and their frequencies and semi-amplitudes are given in Table 5.

We did not include the eclipses in 2000 in this analysis, since the disc was clearly in a state of transition at that time. The frequencies and semi-amplitudes of the six signals marked U to Z in Fig. 14 are listed in Table 5 along with our interpretation of their origin. We also note there is a strong signal in the variation of eclipse skew at the frequency 9.48 c d$^{-1}$ whose origin is unknown.

All three eclipse parameters show strongest signals (marked U) at frequencies close to the nodal precession frequencies seen in the light curve, and also show signals near the positive superhump frequency (V). The orbital signal (X) appears in the eclipse depth and width, while the negative superhump frequency (Y) only appears in the eclipse width. The signal at $\omega$-$N$ (W) is present in all three parameters while the signal $\omega$+$\Omega$ (Z) shows up only in the eclipse depth.

Fig. 15 shows these three eclipse parameters for eclipses in 2014, the year with most data, phased on the period corresponding to the nodal precession frequency listed in Table 3. Fig. 16 shows the same data phased on the apsidal precession frequency in Table 3. Also shown are sinusoidal fits to the data.

**Table 5.** Frequencies and semi-amplitudes of signals from CLEANest analyses of the eclipse depth, width and skew for all sufficiently well observed eclipses between 2001 and 2015 and our interpretation of their origin. The semi-amplitudes of signals seen in eclipse depth are given in magnitudes and of those seen in eclipse width and skew in seconds. Numbers in brackets are uncertainties in the final digit, calculated according to the method in Schwarzenberg-Czerny (1991).

| Signal | | Eclipse depth | Eclipse width | Eclipse skew | Interpretation |
|---|---|---|---|---|---|
| U | Frequency (c d$^{-1}$) | 0.22982(3) | 0.22970(5) | 0.22971(5) | Nodal precession $N$ |
|   | Semi-amplitude | 0.068(20) | 23(11) | 8(4) | |
| V | Frequency (c d$^{-1}$) | 6.90310(4) | 6.92308(6) | 6.90310(7) | +ve superhump $\omega-\Omega$ |
|   | Semi-amplitude | 0.056(20) | 21(11) | 7(4) | |
| W | Frequency (c d$^{-1}$) | 7.10290(4) | 7.10290(6) | 7.10290(8) | $\omega-N$ |
|   | Semi-amplitude | 0.058(20) | 20(11) | 6(4) | |
| X | Frequency (c d$^{-1}$) | 7.32268(5) | 7.32268(6) | | Binary orbit $\omega$ |
|   | Semi-amplitude | 0.039(20) | 21(11) | | |
| Y | Frequency (c d$^{-1}$) | | 7.54246(4) | | -ve superhump $\omega+N$ |
|   | Semi-amplitude | | 28(11) | | |
| Z | Frequency (c d$^{-1}$) | 7.76224(5) | | | $\omega+\Omega$ |
|   | Semi-amplitude | 0.043(20) | | | |

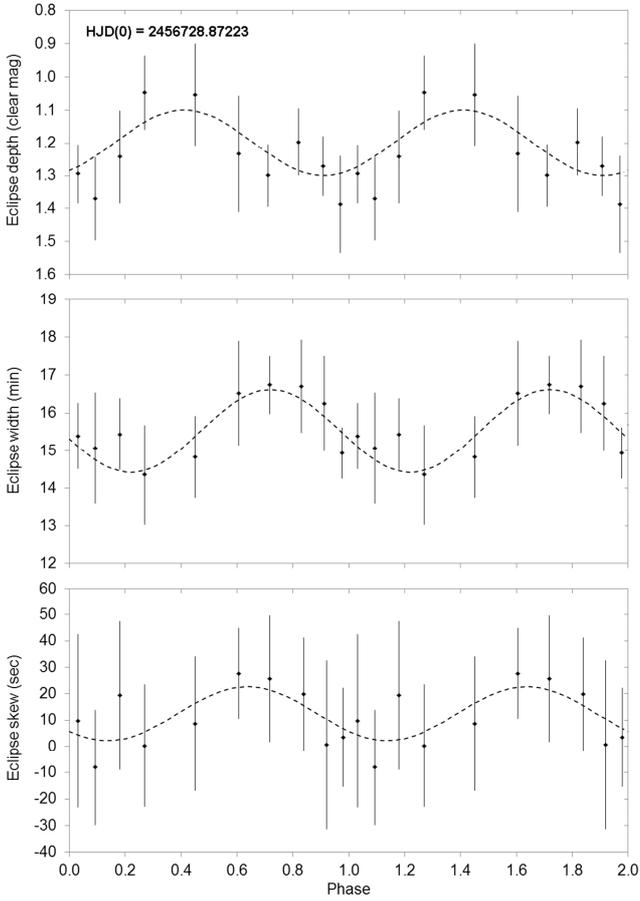

**Figure 15.** Eclipse depth (upper), width (middle) and skew (lower) phased at the period corresponding to the nodal precession frequency in Table 3 with sinusoidal fits to the data. Phase zero corresponds to the maximum magnitude of the nodal precession cycle at the epoch shown. Two cycles are shown and bins are equally populated.

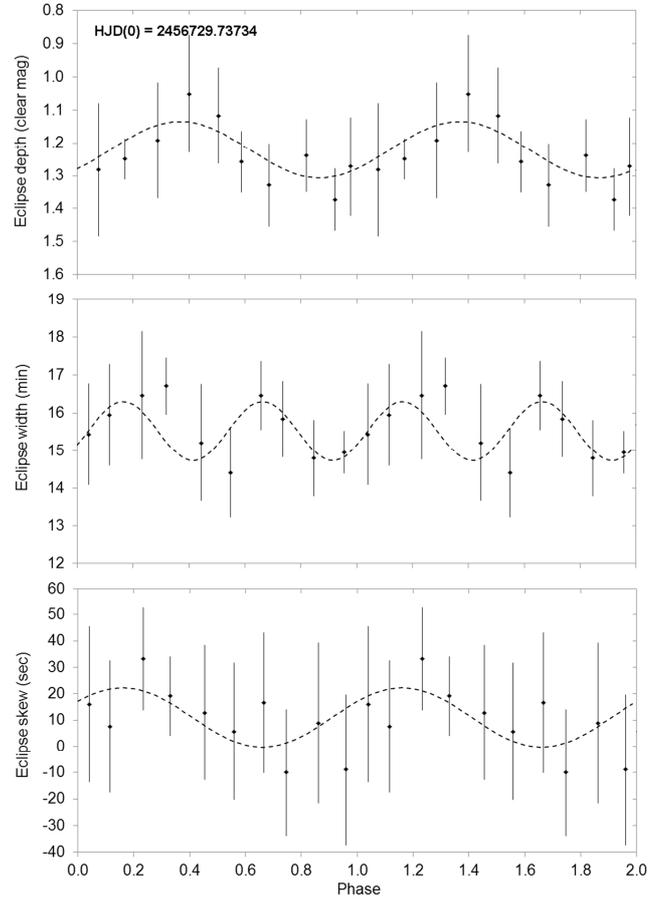

**Figure 16.** Eclipse depth (upper), width (middle) and skew (lower) phased at the period corresponding to the apsidal precession frequency in Table 3 with sinusoidal fits to the data. Phase zero corresponds to the maximum magnitude of the apsidal precession cycle at the epoch shown. Two cycles are shown and bins are equally populated.

In both figures, phase zero corresponds to the maximum magnitude of the appropriate precession cycle at the epoch indicated. In Fig. 16 the sinusoidal fit for the eclipse width with two cycles per phase was statistically better than with one cycle per phase. It is clear from these results that there is indeed a correlation between both nodal and apsidal precession of the accretion disc and eclipse depth, width and skew.

## 7 SIGNALS AT $\omega-N$ AND $\omega+\Omega$

The signals at $\omega-N$ and $\omega+\Omega$, which appear to be persistent over several years, were quite surprising to us. They have not been seen in other stars, and have no straightforward physical interpretation. Could they be a computational artefact? Using simulated data, we attempted to investigate this possibility.

To investigate whether gaps in the light curve caused by removal of the eclipses could give rise to these signals, we generated a sinusoidal signal at the frequency and amplitude of the negative superhump signal and evaluated it at the times of the actual observations outside eclipse in 2014. We then performed a CLEANest analysis on the resulting data. This was repeated for the positive superhump signal. In neither case did this result in a signal being created at either of the above frequencies.

We then generated a sinusoidal negative superhump signal and amplitude-modulated it with an orbital signal, both with their measured frequencies and amplitudes. This was evaluated at the same times as before and subjected to a CLEANest analysis. We found weak signals close to the nodal frequency and at $\omega$-$N$. The exercise was repeated with a positive superhump signal similarly amplitude-modulated with the orbital signal. This produced weak signals close to the apsidal frequency and at $\omega$+$\Omega$.

While it is clear that such modulation can produce signals at the frequencies $\omega$-$N$ and $\omega$+$\Omega$ and also at the nodal and apsidal precession frequencies, the amplitudes of all these signals are 40-50 times smaller than the signals observed in our data. So we are inclined to doubt the hypothesis of an origin in amplitude modulation. But the question is still open. It may be that the absence of $\omega$-$N$ and $\omega$+$\Omega$ signals in other stars is simply due to the sparser coverage available for other stars.

## 8 LOW STATE IN 1999

Our observations in 1999 found DW UMa 3.5 magnitudes below its normal level. Low states of DW UMa have been reported previously (Hessman 1990; Honeycut et al. 1993; Dhillon et al. 1994; Araujo-Betancor et al. 2003) and attributed to a reduction or cessation of mass transfer. It has been suggested that this interruption of mass transfer is due to the presence of one or more magnetically-induced cool starspots on the surface of the MS star at the inner Lagrange point which shut off the accretion stream (Livio & Pringle 1994).

Fig. 17 shows our best defined 1999 eclipse which was recorded with a clear filter on 1999 February 9. The relatively sharp entry into and exit from the eclipse and the steep sides of the eclipse point to the absence of an observable accretion disc at that time. Therefore during the eclipse see only light from the MS star.

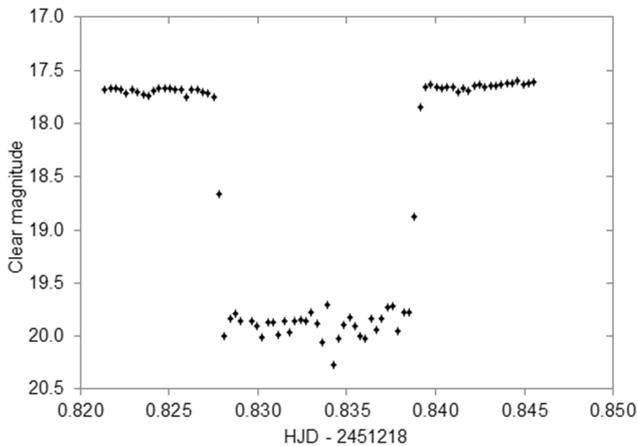

**Figure 17**. Light curve of an eclipse on 1999 February 9 recorded with a clear filter while DW UMa was in a low state.

By linear fits to the out-of-eclipse, ingress, in-eclipse and egress segments of the light curve, we derive an eclipse width at half-flux level of 965±9 s and a mean duration of ingress and egress of 60±9 s. These values are consistent with 969±4 s and 48±3 s respectively reported by Araujo-Betancor et al. (2003) derived from UV observations of eclipses in 1999 January with the HST Imaging Spectrograph. This gives us confidence in adopting their binary system parameters in our analysis.

During the WD eclipse we measured mean $B$, $Rc$ and $Ic$ magnitudes for the MS star as 21.8±0.3, 20.16±0.16 and 18.76±0.13 respectively. Interpolating in Table 2 in Bessell (1991), the ($Rc$-$Ic$) colour index of 1.40±0.21 indicates a spectral type of M3.1±1.0 for the MS star. This compares with M3.5±1.0 derived by Araujo-Betancor et al. (2003) on the basis of their ($I$-$K$) measurement.

Fig. 18 shows $B$, $Rc$ and $Ic$ light curves outside eclipse averaged in 10 orbital phase bins. Measurement uncertainties are smaller than the size of the symbols. These all show a strong maximum at phase 0.5, at which time the MS star is on the far side of the WD. Assuming the apparent brightness of the WD does not change as the binary rotates, this apparent brightening of the MS star by 0.3, 0.6 and 0.8 magnitudes in $B$, $Rc$ and $Ic$ respectively can be explained by irradiation of the side of the cool MS star facing the hot WD. We see this side of the MS star face-on at phase 0.5. Taking the increase in brightness of the MS star at 4500Å as 0.3 magnitudes, this corresponds to a 30 per cent increase in flux. In the absence of an accretion disc, and with the relative sizes and fluxes of the WD and MS star (as reported below), we would not expect to see a secondary eclipse, nor do we see evidence of one.

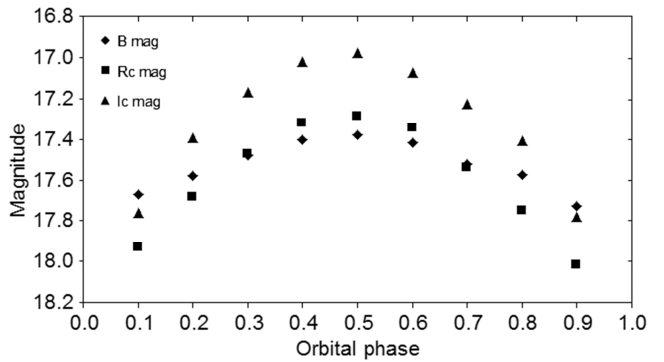

**Figure 18**. $B$, $Rc$ and $Ic$ light curves outside eclipse for 1999 averaged in orbital phase bins of 0.1. Measurement uncertainties are smaller than the symbol sizes.

Biro (2002) reported a similar variation of 0.5 magnitudes in unfiltered light over the orbital cycle in the low state. Dhillon et al. (1994), observed DW UMa spectroscopically during a low state. Comparing their reported continuum flux levels (their fig. 7) with those reported in a high state by Shafter et al. (1988) (their fig. 4) indicates that the magnitude of DW UMa during the observations by Dhillon et al. (1994) was around 18, and we now know this indicates the accretion disc may well have been absent. Dhillon et al. (1994) reported Balmer emission lines originating on the inner face of the MS star which they attributed to irradiation by the WD or accretion disc. The increase of continuum flux between phases 0.1 and 0.5 at 4500Å in Dhillon et al. (1994) (their fig. 7) is consistent with the 30 per cent increase we found photometrically. A similar effect was noted by Rodriguez-Gil et al. (2012) during their observations of the SW Sex star BB Dor in a low state. They observed the presence of Fe

II emission lines from the MS star which peaked at phase 0.5 and which they concluded were caused by irradiation of the MS star by the WD.

## 9 RETURN TO HIGH STATE

Between 1999 and 2002, the accretion disc reformed and DW UMa returned to its normal high state. Fig. 19 shows mean flux profiles of eclipses recorded with a clear filter each year from 1999 to 2002. During the 1999 eclipse, only flux from the MS star was recorded. We assigned this a mean flux level of 1. With the reappearance outside eclipse of the WD the flux increased by a factor of 7.8. In 2002, when DW UMa had returned to its high state and the accretion disc had reformed obscuring the WD behind its rim, the flux outside eclipse was 185 times larger than the 1999 flux during eclipse. Corresponding to their relative flux ratio of 1 : 6.8 : 184, the separate clear magnitudes for the MS star, WD and accretion disc are 19.9, 17.8 and 14.2 confirming that, in its high state, the accretion disc completely dominates the light output of DW UMa. Given these relative magnitudes it is not surprising that we do not see strong evidence of a secondary eclipse in the mean light curves in Fig. 2. The residual flux of approximately 32 percent during eclipse in the high state indicates that a substantial part of the disc remains unobscured. Our observations are consistent with Dhillon et al. (2013) (their fig. 7) which gives a schematic representation of the DW UMa system.

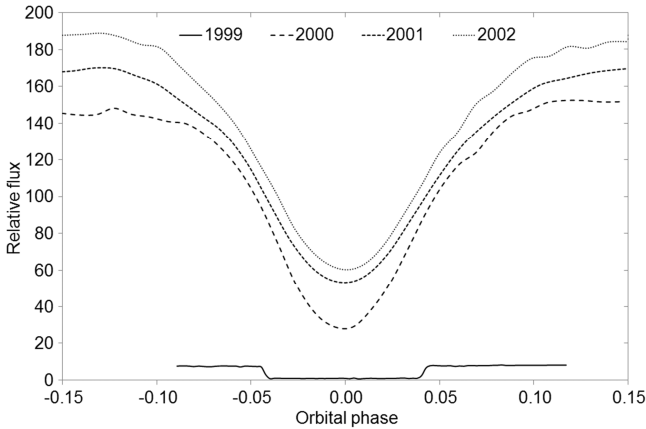

**Figure 19**. Mean flux profiles of eclipses recorded with a clear filter for each year from 1999 to 2002.

By measuring the full eclipse width from start of ingress to end of egress each year from 2000 to 2002, and using the binary system parameters given by Araujo-Betancor et al. (2003), we can estimate the approximate radius of the reforming accretion disc as $0.40\pm0.08$ $R_\odot$ in 2000, $0.51\pm0.07$ $R_\odot$ in 2001 and $0.55\pm0.07$ $R_\odot$ in 2002. This is consistent with an asymptotic disc radius of $0.8L_1$ (= 0.64 $R_\odot$) suggested in Dhillon et al. (1994) (their fig. 4). This is almost twice the radius of the MS star (=0.34 $R_\odot$).

## 10. SUMMARY AND CONCLUSIONS

We have carried out the most extensive study of periodic signals yet obtained for any novalike variable, including the mysterious SW Sex class, for which DW UMa is an excellent prototype. Our principal findings are as follows:

1. Analysis of 372 new and 260 previously published eclipse times, spanning 32 years, shows that the precise orbital period wanders back and forth with a period or quasi-period near 13.6 years. The origin and coherence of this wander are still unknown.

2. We present a detailed account of the comings and goings of superhumps. Positive superhumps were seen in 2002, 2004, 2014 and 2015; and negative superhumps appeared in 2014 and 2015. The corresponding positive and negative superhump period excesses are 0.063 and -0.029. In both 2014 and 2015, we also detected low-frequency signals which we attribute to apsidal and nodal disc precession. The nodal signal was no surprise; this often accompanies negative superhumps (Armstrong et al. 2013). But the apsidal signal, near $\Omega = 0.40$ c/d, was a surprise as no such signal has been reliably seen in the light curves of any other cataclysmic variable.

3. In our two seasons of best coverage, both superhumps were present as well as the precessional signals themselves ($N$ and $\Omega$). This leads us to conjecture that both precessions are always present in the binary, and therefore to the hypothesis that these precessions are key to the SW Sex phenomenon. The wobbling disc could lead to the mass-transfer stream overflowing the disc, and the elliptical disc should shift its Doppler signature away from disc centre. Both may contribute to the "phase shifts" which mysteriously characterize the SW Sex syndrome. Disc wobble can also lead to direct irradiation of the secondary by the hot WD (which would otherwise be shielded by the opening angle of the disc), and that may produce the very high mass-transfer rates.

4. The two types of superhumps may operate quite independently. Within the limits of this study (not ideal, since detection limits were different in different years), neither type seems to affect the presence or properties of the other.

5. We find a long-term progressive change in the apsidal disc precession period, from 2.20 d to 2.55 d between 2002 and 2015.

6. The depth, width and skew of eclipses are all modulated with both nodal and apsidal precession periods. This is expected in principle, since the projected disc shape near eclipse must vary with the precession period(s). We also see a tendency for eclipses to be slightly skewed towards a delayed centre time, which may be due to the bright spot at or near the disc rim.

7. We observe unexpected signals at frequencies $\omega-N$ and $\omega+\Omega$ in the light curve and in the variation of eclipse parameters. We found no entirely plausible explanation for these signals, and could not decisively rule out an origin in amplitude modulation.

8. In 1999 we observed DW UMa in a low state, during which the accretion disc appears to have completely disappeared. In the absence of light from the accretion disc, the photometry suggests a 30% increase in flux from the face of the main sequence star, perhaps because it is irradiated by the hot white dwarf.

9. From the ($Rc$-$Ic$) colour index of 1.40 of the main sequence star during eclipse in the low state, we estimated its spectral type as M3.1±1.0.

10. By measuring how the flux changes between low and high states both in and out of eclipse, we estimate the C (roughly V) relative flux ratio of the main-sequence star, the white dwarf and the fully formed accretion disc to be 1 : 6.8 : 184. Their separate C magnitudes appear to be 19.9, 17.8 and 14.2.

Most significant in this list are the new results on superhumps. Our 2014 data showed how the eclipse shapes varied with the nodal and apsidal precession cycles. This variation provides an incentive for future attempts to model the behaviour of a disc experiencing simultaneous prograde and retrograde precession in the presence of a high accretion rate.

Nearly all SW Sex stars produce superhumps but DW UMa seems to be the most prolific and its deep eclipses provide an extra constraint as the knife-edge of the dark secondary draws across the eccentric and/or wobbling disc. But because our telescopes are small and the eclipse is deep, we cannot fully exploit the information that must be present in the eclipses. We leave this for bigger telescopes. The above results should also motivate future theoretical efforts to study the presence of apsidal and nodal precession, whether alone or combined, in high-inclination discs of high accretion rate.

## ACKNOWLEDGEMENTS

We are grateful to the anonymous referee and to the scientific editor for comments which helped us to improve the paper. We thank the U.S. National Science Foundation for support of this research through grants AST 1211129 and 1615456 to Columbia University. We acknowledge the work of the AAVSO in providing comparison star charts and sequences. This research has made use of the SIMBAD database, operated at CDS, Strasbourg, France.

## APPENDIX: ECLIPSE TIMINGS

**Table 2.** Times of minimum, scaled errors and cycle numbers of 372 measured eclipses.

| HJD | Error | Cycle | HJD | Error | Cycle | HJD | Error | Cycle |
|---|---|---|---|---|---|---|---|---|
| 2451210.77362 | 0.00040 | 0 | 2451627.83293 | 0.00050 | 3053 | 2451633.57071 | 0.00024 | 3095 |
| 2451217.05719 | 0.00060 | 46 | 2451628.65257 | 0.00031 | 3059 | 2451987.79103 | 0.00014 | 5688 |
| 2451218.01388 | 0.00040 | 53 | 2451628.65266 | 0.00036 | 3059 | 2451988.74751 | 0.00025 | 5695 |
| 2451218.83332 | 0.00040 | 59 | 2451628.78951 | 0.00027 | 3060 | 2451993.80150 | 0.00031 | 5732 |
| 2451218.97010 | 0.00060 | 60 | 2451629.74614 | 0.00068 | 3067 | 2451996.80706 | 0.00018 | 5754 |
| 2451219.78935 | 0.00060 | 66 | 2451629.74616 | 0.00018 | 3067 | 2452001.45168 | 0.00063 | 5788 |
| 2451605.97628 | 0.00015 | 2893 | 2451629.88197 | 0.00017 | 3068 | 2452014.56635 | 0.00025 | 5884 |
| 2451617.86095 | 0.00018 | 2980 | 2451630.56480 | 0.00017 | 3073 | 2452015.65962 | 0.00074 | 5892 |
| 2451617.99753 | 0.00014 | 2981 | 2451630.70164 | 0.00038 | 3074 | 2452015.79558 | 0.00054 | 5893 |
| 2451618.68058 | 0.00015 | 2986 | 2451630.70176 | 0.00012 | 3074 | 2452022.62632 | 0.00051 | 5943 |
| 2451618.81671 | 0.00016 | 2987 | 2451630.70198 | 0.00071 | 3074 | 2452023.71825 | 0.00037 | 5951 |
| 2451620.72945 | 0.00025 | 3001 | 2451630.83860 | 0.00016 | 3075 | 2452024.67540 | 0.00029 | 5958 |
| 2451620.86602 | 0.00031 | 3002 | 2451632.61424 | 0.00017 | 3088 | 2452024.81225 | 0.00046 | 5959 |
| 2451621.00286 | 0.00038 | 3003 | 2451632.75100 | 0.00083 | 3089 | 2452025.63133 | 0.00010 | 5965 |
| 2451627.69655 | 0.00033 | 3052 | 2451632.75111 | 0.00015 | 3089 | 2452025.76778 | 0.00044 | 5966 |

| HJD | Error | Cycle | HJD | Error | Cycle | HJD | Error | Cycle |
|---|---|---|---|---|---|---|---|---|
| 2452028.63742 | 0.00030 | 5987 | 2453100.86178 | 0.00039 | 13836 | 2456015.63518 | 0.00026 | 35173 |
| 2452029.59301 | 0.00030 | 5994 | 2453100.86181 | 0.00040 | 13836 | 2456029.43239 | 0.00019 | 35274 |
| 2452030.68614 | 0.00021 | 6002 | 2453101.54498 | 0.00059 | 13841 | 2456033.39481 | 0.00017 | 35303 |
| 2452031.64185 | 0.00025 | 6009 | 2453101.68108 | 0.00030 | 13842 | 2456088.44605 | 0.00018 | 35706 |
| 2452031.77902 | 0.00022 | 6010 | 2453101.68129 | 0.00039 | 13842 | 2456366.44132 | 0.00028 | 37741 |
| 2452036.69673 | 0.00044 | 6046 | 2453101.81715 | 0.00034 | 13843 | 2456366.57694 | 0.00020 | 37742 |
| 2452037.65342 | 0.00033 | 6053 | 2453105.77984 | 0.00016 | 13872 | 2456382.42441 | 0.00022 | 37858 |
| 2452039.56529 | 0.00054 | 6067 | 2453105.91647 | 0.00026 | 13873 | 2456384.47293 | 0.00011 | 37873 |
| 2452040.52148 | 0.00048 | 6074 | 2453106.59901 | 0.00059 | 13878 | 2456399.36316 | 0.00032 | 37982 |
| 2452042.43396 | 0.00049 | 6088 | 2453106.73542 | 0.00024 | 13879 | 2456407.42226 | 0.00016 | 38041 |
| 2452042.57061 | 0.00048 | 6089 | 2453108.64822 | 0.00062 | 13893 | 2456408.37935 | 0.00024 | 38048 |
| 2452311.68591 | 0.00023 | 8059 | 2453108.78469 | 0.00055 | 13894 | 2456413.43337 | 0.00030 | 38085 |
| 2452311.82219 | 0.00014 | 8060 | 2453109.60450 | 0.00037 | 13900 | 2456413.43356 | 0.00018 | 38085 |
| 2452311.95895 | 0.00034 | 8061 | 2453109.74127 | 0.00046 | 13901 | 2456728.44827 | 0.00013 | 40391 |
| 2452312.77848 | 0.00015 | 8067 | 2453110.69719 | 0.00045 | 13908 | 2456732.40957 | 0.00036 | 40420 |
| 2452312.91517 | 0.00012 | 8068 | 2453111.51740 | 0.00049 | 13914 | 2456732.54516 | 0.00031 | 40421 |
| 2452312.91527 | 0.00012 | 8068 | 2453111.65365 | 0.00046 | 13915 | 2456732.68314 | 0.00036 | 40422 |
| 2452313.05187 | 0.00017 | 8069 | 2453111.65377 | 0.00052 | 13915 | 2456733.36651 | 0.00034 | 40427 |
| 2452319.74504 | 0.00019 | 8118 | 2453111.79016 | 0.00025 | 13916 | 2456733.50233 | 0.00014 | 40428 |
| 2452319.88117 | 0.00020 | 8119 | 2453111.79046 | 0.00031 | 13916 | 2456733.63929 | 0.00032 | 40429 |
| 2452320.56460 | 0.00049 | 8124 | 2453112.61002 | 0.00023 | 13922 | 2456733.63930 | 0.00032 | 40429 |
| 2452320.70182 | 0.00023 | 8125 | 2453112.74624 | 0.00018 | 13923 | 2456733.77583 | 0.00028 | 40430 |
| 2452320.70189 | 0.00024 | 8125 | 2453112.74630 | 0.00032 | 13923 | 2456734.45811 | 0.00033 | 40435 |
| 2452320.83784 | 0.00025 | 8126 | 2454181.41972 | 0.00012 | 21746 | 2456734.59495 | 0.00031 | 40436 |
| 2452320.83826 | 0.00030 | 8126 | 2454185.38111 | 0.00023 | 21775 | 2456734.59504 | 0.00020 | 40436 |
| 2452321.38476 | 0.00028 | 8130 | 2454224.45069 | 0.00028 | 22061 | 2456734.73206 | 0.00068 | 40437 |
| 2452321.52156 | 0.00043 | 8131 | 2454473.34784 | 0.00028 | 23883 | 2456735.41477 | 0.00026 | 40442 |
| 2452321.65807 | 0.00029 | 8132 | 2454564.46472 | 0.00026 | 24550 | 2456735.55158 | 0.00033 | 40443 |
| 2452322.34110 | 0.00033 | 8137 | 2454580.44810 | 0.00028 | 24667 | 2456735.68738 | 0.00043 | 40444 |
| 2452322.61451 | 0.00047 | 8139 | 2454580.58430 | 0.00017 | 24668 | 2456737.46334 | 0.00020 | 40457 |
| 2452323.57065 | 0.00036 | 8146 | 2454588.37104 | 0.00024 | 24725 | 2456737.60081 | 0.00025 | 40458 |
| 2452323.70699 | 0.00025 | 8147 | 2454588.50719 | 0.00014 | 24726 | 2456738.42056 | 0.00031 | 40464 |
| 2452324.39053 | 0.00057 | 8152 | 2454593.42488 | 0.00018 | 24762 | 2456738.55694 | 0.00029 | 40465 |
| 2452324.66326 | 0.00023 | 8154 | 2454596.43088 | 0.00033 | 24784 | 2456739.37654 | 0.00014 | 40471 |
| 2452327.80551 | 0.00021 | 8177 | 2454884.39718 | 0.00020 | 26892 | 2456740.60560 | 0.00025 | 40480 |
| 2452327.94205 | 0.00049 | 8178 | 2454892.32022 | 0.00021 | 26950 | 2456742.38226 | 0.00019 | 40493 |
| 2452329.44417 | 0.00071 | 8189 | 2455239.30027 | 0.00017 | 29490 | 2456742.65496 | 0.00018 | 40495 |
| 2452330.67418 | 0.00018 | 8198 | 2455263.34322 | 0.00012 | 29666 | 2456742.65511 | 0.00034 | 40495 |
| 2452330.81015 | 0.00017 | 8199 | 2455270.30998 | 0.00013 | 29717 | 2456742.79126 | 0.00018 | 40496 |
| 2452330.81077 | 0.00041 | 8199 | 2455278.37037 | 0.00014 | 29776 | 2456742.79181 | 0.00031 | 40496 |
| 2452330.94743 | 0.00035 | 8200 | 2455624.94125 | 0.00022 | 32313 | 2456742.92792 | 0.00018 | 40497 |
| 2452331.76637 | 0.00033 | 8206 | 2455627.39964 | 0.00024 | 32331 | 2456743.47452 | 0.00021 | 40501 |
| 2452331.90325 | 0.00033 | 8207 | 2455627.39979 | 0.00014 | 32331 | 2456743.61080 | 0.00025 | 40502 |
| 2452332.03913 | 0.00046 | 8208 | 2455627.53661 | 0.00030 | 32332 | 2456745.66115 | 0.00021 | 40517 |
| 2453092.39230 | 0.00023 | 13774 | 2455627.67295 | 0.00016 | 32333 | 2456745.79740 | 0.00016 | 40518 |
| 2453092.52901 | 0.00019 | 13775 | 2455628.35618 | 0.00014 | 32338 | 2456745.93387 | 0.00015 | 40519 |
| 2453092.66507 | 0.00022 | 13776 | 2455629.31207 | 0.00031 | 32345 | 2456746.34402 | 0.00025 | 40522 |
| 2453092.66529 | 0.00021 | 13776 | 2455688.59904 | 0.00015 | 32779 | 2456746.48049 | 0.00025 | 40523 |
| 2453092.80215 | 0.00026 | 13777 | 2455691.60442 | 0.00028 | 32801 | 2456746.61663 | 0.00055 | 40524 |
| 2453092.80219 | 0.00027 | 13777 | 2455692.69843 | 0.00029 | 32809 | 2456746.61664 | 0.00037 | 40524 |
| 2453093.62193 | 0.00035 | 13783 | 2455983.66857 | 0.00043 | 34939 | 2456746.75391 | 0.00031 | 40525 |
| 2453093.62207 | 0.00016 | 13783 | 2455983.80540 | 0.00045 | 34940 | 2456747.70952 | 0.00023 | 40532 |
| 2453093.75823 | 0.00027 | 13784 | 2455984.62557 | 0.00041 | 34946 | 2456747.84565 | 0.00031 | 40533 |
| 2453093.75828 | 0.00027 | 13784 | 2455984.62567 | 0.00028 | 34946 | 2456750.57849 | 0.00060 | 40553 |
| 2453093.75849 | 0.00028 | 13784 | 2455984.76203 | 0.00026 | 34947 | 2456751.39859 | 0.00022 | 40559 |
| 2453093.89508 | 0.00032 | 13785 | 2455985.58190 | 0.00032 | 34953 | 2456751.53446 | 0.00022 | 40560 |
| 2453093.89516 | 0.00035 | 13785 | 2455985.71830 | 0.00035 | 34954 | 2456751.67102 | 0.00025 | 40561 |
| 2453094.57756 | 0.00036 | 13790 | 2455985.85487 | 0.00026 | 34955 | 2456751.67123 | 0.00027 | 40561 |
| 2453095.67019 | 0.00037 | 13798 | 2455986.81076 | 0.00015 | 34962 | 2456752.76397 | 0.00021 | 40569 |
| 2453095.67074 | 0.00015 | 13798 | 2455987.90418 | 0.00018 | 34970 | 2456753.44745 | 0.00020 | 40574 |
| 2453095.80684 | 0.00046 | 13799 | 2455989.95362 | 0.00026 | 34985 | 2456753.58271 | 0.00022 | 40575 |
| 2453095.80724 | 0.00035 | 13799 | 2455990.90946 | 0.00049 | 34992 | 2456753.58289 | 0.00048 | 40575 |
| 2453096.62701 | 0.00069 | 13805 | 2455991.45632 | 0.00023 | 34996 | 2456753.58293 | 0.00020 | 40575 |
| 2453096.62784 | 0.00056 | 13805 | 2455998.69635 | 0.00062 | 35049 | 2456753.72013 | 0.00014 | 40576 |
| 2453096.76394 | 0.00014 | 13806 | 2455998.83265 | 0.00024 | 35050 | 2456753.72033 | 0.00024 | 40576 |
| 2453096.76399 | 0.00040 | 13806 | 2455998.96947 | 0.00015 | 35051 | 2456754.40324 | 0.00012 | 40581 |
| 2453097.58337 | 0.00072 | 13812 | 2456000.60880 | 0.00025 | 35063 | 2456754.53990 | 0.00063 | 40582 |
| 2453097.71967 | 0.00040 | 13813 | 2456001.56422 | 0.00089 | 35070 | 2456754.54000 | 0.00022 | 40582 |
| 2453097.71990 | 0.00098 | 13813 | 2456002.65766 | 0.00037 | 35078 | 2456754.54005 | 0.00023 | 40582 |
| 2453097.85677 | 0.00092 | 13814 | 2456002.79427 | 0.00012 | 35079 | 2456754.81353 | 0.00017 | 40584 |
| 2453099.76840 | 0.00033 | 13828 | 2456002.93060 | 0.00041 | 35080 | 2456754.95000 | 0.00017 | 40585 |
| 2453100.72506 | 0.00078 | 13835 | 2456003.61394 | 0.00027 | 35085 | 2456755.35973 | 0.00034 | 40588 |

| HJD | Error | Cycle | HJD | Error | Cycle | HJD | Error | Cycle |
|---|---|---|---|---|---|---|---|---|
| 2456755.49629 | 0.00023 | 40589 | 2456761.91689 | 0.00018 | 40636 | 2457089.36247 | 0.00018 | 43033 |
| 2456755.49631 | 0.00021 | 40589 | 2456762.46266 | 0.00024 | 40640 | 2457089.49945 | 0.00037 | 43034 |
| 2456755.63283 | 0.00015 | 40590 | 2456763.41896 | 0.00023 | 40647 | 2457089.63561 | 0.00029 | 43035 |
| 2456755.76902 | 0.00021 | 40591 | 2456763.55603 | 0.00039 | 40648 | 2457090.45484 | 0.00035 | 43041 |
| 2456755.90603 | 0.00046 | 40592 | 2456763.69264 | 0.00028 | 40649 | 2457091.41172 | 0.00025 | 43048 |
| 2456756.45225 | 0.00028 | 40596 | 2456764.37561 | 0.00030 | 40654 | 2457091.54829 | 0.00030 | 43049 |
| 2456756.58889 | 0.00025 | 40597 | 2456764.51222 | 0.00031 | 40655 | 2457092.36806 | 0.00014 | 43055 |
| 2456756.72586 | 0.00050 | 40598 | 2456764.64861 | 0.00031 | 40656 | 2457092.64143 | 0.00017 | 43057 |
| 2456756.72596 | 0.00035 | 40598 | 2456764.64884 | 0.00029 | 40656 | 2457093.32400 | 0.00022 | 43062 |
| 2456756.72605 | 0.00037 | 40598 | 2456765.46825 | 0.00034 | 40662 | 2457093.46050 | 0.00021 | 43063 |
| 2456756.86199 | 0.00030 | 40599 | 2456765.60525 | 0.00036 | 40663 | 2457093.59662 | 0.00032 | 43064 |
| 2456757.40911 | 0.00020 | 40603 | 2456765.74148 | 0.00022 | 40664 | 2457094.41670 | 0.00027 | 43070 |
| 2456757.68130 | 0.00032 | 40605 | 2456766.42470 | 0.00016 | 40669 | 2457094.55302 | 0.00047 | 43071 |
| 2456757.68138 | 0.00016 | 40605 | 2456766.56143 | 0.00020 | 40670 | 2457095.37388 | 0.00022 | 43077 |
| 2456757.68147 | 0.00029 | 40605 | 2456767.65387 | 0.00034 | 40678 | 2457095.50996 | 0.00019 | 43078 |
| 2456757.81897 | 0.00030 | 40606 | 2456768.60969 | 0.00036 | 40685 | 2457095.64650 | 0.00021 | 43079 |
| 2456758.36403 | 0.00040 | 40610 | 2456768.60971 | 0.00026 | 40685 | 2457096.32910 | 0.00020 | 43084 |
| 2456758.63897 | 0.00024 | 40612 | 2457020.37616 | 0.00021 | 42528 | 2457096.46618 | 0.00028 | 43085 |
| 2456758.77523 | 0.00038 | 40613 | 2457021.46906 | 0.00017 | 42536 | 2457096.60238 | 0.00020 | 43086 |
| 2456758.91145 | 0.00031 | 40614 | 2457075.42848 | 0.00020 | 42931 | 2457096.60254 | 0.00019 | 43086 |
| 2456759.45729 | 0.00036 | 40618 | 2457078.43386 | 0.00022 | 42953 | 2457096.73905 | 0.00031 | 43087 |
| 2456759.45754 | 0.00027 | 40618 | 2457078.57086 | 0.00012 | 42954 | 2457097.42228 | 0.00030 | 43092 |
| 2456759.59468 | 0.00059 | 40619 | 2457078.70745 | 0.00023 | 42955 | 2457101.52026 | 0.00016 | 43122 |
| 2456760.41397 | 0.00020 | 40625 | 2457079.39014 | 0.00022 | 42960 | 2457101.65731 | 0.00020 | 43123 |
| 2456760.41406 | 0.00048 | 40625 | 2457079.52668 | 0.00030 | 42961 | 2457102.61299 | 0.00044 | 43130 |
| 2456760.41408 | 0.00025 | 40625 | 2457079.66311 | 0.00025 | 42962 | 2457106.43837 | 0.00012 | 43158 |
| 2456760.55016 | 0.00057 | 40626 | 2457080.48276 | 0.00016 | 42968 | 2457107.39498 | 0.00076 | 43165 |
| 2456760.55075 | 0.00035 | 40626 | 2457081.84951 | 0.00046 | 42978 | 2457108.35064 | 0.00012 | 43172 |
| 2456760.68726 | 0.00026 | 40627 | 2457082.66862 | 0.00019 | 42984 | 2457108.48687 | 0.00021 | 43173 |
| 2456760.82343 | 0.00032 | 40628 | 2457082.80547 | 0.00018 | 42985 | 2457109.71713 | 0.00013 | 43182 |
| 2456760.96017 | 0.00031 | 40629 | 2457082.94209 | 0.00020 | 42986 | 2457110.39973 | 0.00017 | 43187 |
| 2456761.36981 | 0.00024 | 40632 | 2457083.07885 | 0.00028 | 42987 | 2457110.53675 | 0.00020 | 43188 |
| 2456761.36987 | 0.00033 | 40632 | 2457083.35201 | 0.00029 | 42989 | 2457110.67374 | 0.00028 | 43189 |
| 2456761.50619 | 0.00029 | 40633 | 2457083.48869 | 0.00027 | 42990 | 2457110.67425 | 0.00045 | 43189 |
| 2456761.50632 | 0.00031 | 40633 | 2457083.62533 | 0.00033 | 42991 | 2457110.81008 | 0.00016 | 43190 |
| 2456761.50639 | 0.00028 | 40633 | 2457084.44530 | 0.00042 | 42997 | | | |
| 2456761.64320 | 0.00040 | 40634 | 2457084.58164 | 0.00016 | 42998 | | | |
| 2456761.78107 | 0.00023 | 40635 | 2457084.71854 | 0.00023 | 42999 | | | |